# Experimental Quantum Teleportation of a Two-Qubit Composite System

Qiang Zhang[2,1], Alexander Goebel[1], Claudia Wagenknecht[1], Yu-Ao Chen[1], Bo Zhao[1], Tao Yang[2], Alois Mair[1], Jörg Schmiedmayer[1,3] & Jian-Wei Pan[1,2]∗

[1] Physikalisches Institut, Universität Heidelberg, Philosophenweg 12, D-69120 Heidelberg, Germany
[2] Hefei National Laboratory for Physical Sciences at Microscale and Department of Modern Physics, University of Science and Technology of China, Hefei, Anhui, 230027, PR China
[3] Atominstitut der Österreichischen Universitäten, TU-Wien, A-1020 Vienna, Austria

**Quantum teleportation[1], a way to transfer the state of a quantum system from one location to another, is central to quantum communication[2] and plays an important role in a number of quantum computation protocols[3-5]. Previous experimental demonstrations have been implemented with photonic[6-8] or ionic qubits[9,10]. Very recently long-distance teleportation[11,12] and open-destination teleportation[13] have also been realized. Until now, previous experiments[6-13] have only been able to teleport single qubits. However, since teleportation of single qubits is insufficient for a large-scale realization of quantum communication and computation[2-5], teleportation of a composite system containing two or more qubits has been seen as a long-standing goal in quantum information science. Here, we present the experimental realization of quantum teleportation of a two-qubit composite system. In the experiment, we develop and exploit a six-photon interferometer to teleport an arbitrary polarization state of two photons. The observed teleportation fidelities for different initial states are all**

**well beyond the state estimation limit of 0.40 for a two-qubit system[14]. Not only does our six-photon interferometer provide an important step towards teleportation of a complex system, it will also enable future experimental investigations on a number of fundamental quantum communication and computation protocols such as multi-stage realization of quantum-relay[15], fault-tolerant quantum computation[3], universal quantum error-correction[16] and one-way quantum computation[17,18].**

Our experimental scheme[19,20] closely follows the original proposal for teleportation of single qubits[1]. In the two-qubit teleportation, the sender, Alice, wants to send an unknown state of a system composed of qubits 1 and 2,

$$|\chi\rangle_{12} = \alpha|H\rangle_1|H\rangle_2 + \beta|H\rangle_1|V\rangle_2 + \gamma|V\rangle_1|H\rangle_2 + \delta|V\rangle_1|V\rangle_2, \quad (1)$$

to a distant receiver, Bob (Fig. 1). Here we discuss our protocol in terms of polarization photonic qubits. $|H\rangle$ and $|V\rangle$ denote horizontal and vertical linear polarizations respectively; while $\alpha$, $\beta$, $\gamma$ and $\delta$ are four arbitrary complex numbers satisfying $|\alpha|^2 + |\beta|^2 + |\gamma|^2 + |\delta|^2 = 1$. To achieve teleportation, Alice and Bob first have to share two ancillary entangled photon pairs (photon pairs 3-5 and 4-6) which are prepared in the Bell-state $|\Phi^+\rangle = (|H\rangle|H\rangle + |V\rangle|V\rangle)/\sqrt{2}$. The two-qubit teleportation scheme then works as follows:

Alice first teleports the state of photon 1 to photon 5 following the standard teleportation protocol. In terms of the four Bell-states of photons 1 and 3, $|\Phi^\pm\rangle_{13} = (|H\rangle_1|H\rangle_3 \pm |V\rangle_1|V\rangle_3)/\sqrt{2}$ and $|\Psi^\pm\rangle_{13} = (|H\rangle_1|V\rangle_3 \pm |V\rangle_1|H\rangle_3)/\sqrt{2}$, the combined state of photons 1, 2, 3 and 5 can be rewritten as

$$|\chi\rangle_{12}|\Phi^+\rangle_{35} = \frac{1}{2}(|\Phi^+\rangle_{13}|\chi\rangle_{52} + |\Phi^-\rangle_{13}\hat{\sigma}_{5z}|\chi\rangle_{52} + |\Psi^+\rangle_{13}\hat{\sigma}_{5x}|\chi\rangle_{52} + |\Psi^-\rangle_{13}(-i\hat{\sigma}_{5y}|\chi\rangle_{52})), \quad (2)$$

where $\hat{\sigma}_x$, $\hat{\sigma}_y$ and $\hat{\sigma}_z$ are the well-known Pauli operators. Equation (2) implies, that by performing a joint Bell-state measurement (BSM) on qubits 1 and 3, Alice projects the state of qubits 5 and 2 onto one of the four corresponding states. After she has told Bob her BSM result via a classical communication channel, Bob can convert the state of qubits 5 and 2 into the original state $|\chi\rangle_{52}$ by applying to photon 5 a corresponding local unitary transformation ($\hat{I}, \hat{\sigma}_x, \hat{\sigma}_y, \hat{\sigma}_z$), independent of the original state.

Similarly, the combined state of photons 2, 4, 5 and 6 can be rewritten in terms of the four Bell-states of photons 2 and 4 as

$$|\chi\rangle_{52}|\Phi^+\rangle_{46} = \frac{1}{2}(|\Phi^+\rangle_{24}|\chi\rangle_{56} + |\Phi^-\rangle_{24}\hat{\sigma}_{6z}|\chi\rangle_{56} + |\Psi^+\rangle_{24}\hat{\sigma}_{6x}|\chi\rangle_{56} + |\Psi^-\rangle_{24}(-i\hat{\sigma}_{6y}|\chi\rangle_{56})). \quad (3)$$

Following the above procedure, Alice can also teleport the state of photon 2 to photon 6. First, Alice performs a joint BSM on photons 2 and 4 and sends the BSM result to Bob. Upon the BSM result received, by applying to photon 6 a corresponding local unitary transformation ($\hat{I}, \hat{\sigma}_x, \hat{\sigma}_y, \hat{\sigma}_z$), Bob can convert the state of qubits 5 and 6 into the original state

$$|\chi\rangle_{56} = \alpha|H\rangle_5|H\rangle_6 + \beta|H\rangle_5|V\rangle_6 + \gamma|V\rangle_5|H\rangle_6 + \delta|V\rangle_5|V\rangle_6 \quad (4)$$

to accomplish the task of the most general two-qubit teleportation.

The above scheme has a remarkable feature, that is, it teleports the two photonic qubits, 1 and 2, individually. This way, neither the two original qubits nor the teleported qubits have to be in the same place. Such a flexibility is desired in distributed quantum information processing, such as quantum telecomputation[5] and quantum secret sharing[21]. Moreover, the way of teleporting each qubit of a composite system individually can be easily generalized to teleport a N-qubit complex system.

Although significant experimental advances have been achieved in teleportation of single qubits (photons and ions), the realization of teleportation of a composite system containing two or more qubits has remained a real experimental challenge. This is because, on one hand the recent photonic experiments[11-13] had a too low six-photon coincidence rate. On the other hand, the experiments with trapped ions[9,10] are limited by the finite life time of ion qubits due to decoherence and the nonideal fidelity of quantum logic operation between ion qubits. Since photons are robust against decoherence and high precision unitary transformations for photons can be performed with linear optical devices, in the present experiment we still chose to use polarization-entangled photon pairs via parametric down-conversion[22] as the main resource while various efforts have been made to greatly improve the brightness and stability of the entangled photon sources.

A schematic drawing of our experimental set-up is shown in Fig. 2. A high-intensity UV laser (see Supplementary Information) successively passes through two BBO crystals to generate three polarization entangled photon pairs[22]. The UV laser with a central wavelength of 390nm has a pulse duration of 180fs, a repetition rate of 76MHz and an average power of 1.0W. All three photon pairs are originally prepared in the Bell-state $|\Phi^+\rangle = (|H\rangle|H\rangle + |V\rangle|V\rangle)/\sqrt{2}$. Besides the high intensity UV laser, significant efforts have been made to achieve better collection efficiency and stability of the entangled photon sources (see Supplementary Information). After these efforts, we managed to observe on average $10^5$ photon pairs per second from each source. This is almost 5 times brighter than the source achieved in a recent teleportation experiment[13]. With this high-intensity entangled photon source we could obtain in total 10 six-photon events per minute. This is two orders of magnitude higher than any former photonic teleportation experiments could have achieved.

In the experiment, with the help of wave plates and polarizers inserted in modes 1 and 2, photon pair 1-2 is prepared in the original two-qubit state $|\chi\rangle_{12}$ that is to be teleported. Photon pairs 3-5 and 4-6, which are in the state $|\Phi^+\rangle$, are used as the two ancillary pairs.

To implement two-qubit teleportation, it is necessary to perform a joint BSM on photons 1 and 3 and photons 2 and 4, respectively. Obviously, to demonstrate the working principle of two-qubit teleportation it is sufficient to identify one of the four Bell-states in both BSMs, although this will result in a reduced efficiency—the fraction of success—of 1/16. In the experiment, we decide to analyze the Bell-state $|\Phi^+\rangle$. This is achieved by interfering photons 1 and 3 and photons 2 and 4 on a polarizing beam-splitter, PBS13 and PBS24, respectively. In order to interfere photons 1 and 3 (photons 2 and 4) on the PBS13 (PBS24), one has to guarantee that the two photons have good spatial and temporal overlap at the PBS such that they are indistinguishable. To achieve this, the two outputs of the PBSs are spectrally filtered ( $\Delta\lambda_{FWHM} = 2.8 nm$ ) and monitored by fiber-coupled single-photon detectors[23]. Moreover, perfect temporal overlap is accomplished by adjusting the path length of photon 3 (2) by a delay prism 1 (prism 2) to observe "Hong-Ou-Mandel"-type interference fringes behind the PBS13 (PBS24) in the $+/-$ basis[24], where $|\pm\rangle = (|H\rangle \pm |V\rangle)/\sqrt{2}$. The required projection of photons 1 and 3 (2 and 4) onto $|\Phi^+\rangle$ can thus be achieved by detecting behind PBS13 (PBS24) a $|+\rangle|+\rangle$ or $|-\rangle|-\rangle$ coincidence between detectors D1 and D3 (D2 and D 4)[24]. Note that, in the experiment only the $|+\rangle|+\rangle$ coincidence is registered, which further reduces the teleportation efficiency to 1/64. However, using more detectors one can increase the efficiency up to $1/4$[24].

As shown in equations (2) and (3), the projection measurements onto $|\Phi^+\rangle_{13}$ and $|\Phi^+\rangle_{24}$ leave photons 5 and 6 in the state $|\chi\rangle_{56}$, i.e. the original state of photons 1 and 2. To demonstrate that our two-qubit teleportation protocol works for a general unknown polarization state of photons 1 and 2, we decide to teleport three different initial states $|\chi\rangle_A = |H\rangle|V\rangle$, $|\chi\rangle_C = (|H\rangle + |V\rangle)(|H\rangle - i|V\rangle)/\sqrt{2}$ and $|\chi\rangle_C = (|H\rangle|V\rangle - |V\rangle|H\rangle)/\sqrt{2}$. $|\chi\rangle_A$ is simply one of the four computational basis vectors in the two-qubit Bloch sphere; $|\chi\rangle_B$ is composed by a linear polarization state and a circular polarization state, which is also a superposition of all the four computational basis vectors; while $|\chi\rangle_C$ is a maximally entangled state.

We quantify the quality of our teleportation experiment by looking at the fidelity as defined by

$$F = Tr(\hat{\rho}|\chi\rangle\langle\chi|), \qquad (5)$$

where $|\chi\rangle$ is the original state and $\hat{\rho}$ is the density matrix of the teleported state. To measure the fidelity of two-qubit teleportation, two PBSs (PBS5 and PBS6) and corresponding wave plates (HWP and QWP), as shown in Fig. 2, are combined properly to analyze the teleported state of photons 5 and 6.

The fidelity measurements for the $|\chi\rangle_A$ and $|\chi\rangle_B$ teleportation are straight forward. Conditioned on detecting a $|+\rangle|+\rangle$ coincidence between D1 and D3, D2 and D4, respectively, we analyze the teleported state of photons 5 and 6 in the $H/V$ basis for the $|\chi\rangle_A$ teleportation; while we analyze photon 5 in the $+/-$ basis and photon 6 in the $L/R$ basis for the $|\chi\rangle_B$ teleportation, where $|L\rangle = (|H\rangle + i|V\rangle)/\sqrt{2}$ and $|R\rangle = (|H\rangle - i|V\rangle)/\sqrt{2}$ correspond to the left and right circular polarization states. Since the above state analysis only involves orthogonal measurements on individual qubits, the fidelity of the teleported state is directly given by the fraction of observing a $|\chi\rangle_A$ or $|\chi\rangle_B$ state at detectors D5 and D6. The measurement results are shown in Fig. 3. The experimental integration time for each fidelity measurement was about 60 hours and we recorded about 100 desired two-qubit teleportation events. The integration time is slightly longer than one would expect from the original source rate, due to the additional losses at the interference PBSs. With these original datas, we can conclude that the fidelity for $|\chi\rangle_A$ or $|\chi\rangle_B$ is $0.86 \pm 0.03$ or $0.75 \pm 0.02$ respectively.

The measurement on the fidelity of the $|\chi\rangle_C$ teleportation is a bit more complex, since a complete Bell-state analysis on photons 5 and 6 usually requires nonlinear interaction between them. Fortunately, the fidelity can still be determined by local measurements on individual qubits. To see this, we write the density matrix of $|\chi\rangle_C$ in terms of the Pauli matrices:

$$|\chi\rangle_C \langle\chi| = |\Psi^-\rangle\langle\Psi^-| = \frac{1}{4}(\hat{I} - \hat{\sigma}_x\hat{\sigma}_x - \hat{\sigma}_y\hat{\sigma}_y - \hat{\sigma}_z\hat{\sigma}_z). \qquad (6)$$

By equation (5), we thus have:

$$F = Tr(\hat{\rho}|\psi^-\rangle\langle\psi^-|) = \frac{1}{4}Tr[\hat{\rho}(\hat{I} - \hat{\sigma}_x\hat{\sigma}_x - \hat{\sigma}_y\hat{\sigma}_y - \hat{\sigma}_z\hat{\sigma}_z)]. \qquad (7)$$

This implies that, we can obtain the fidelity of the teleported state by only performing three local measurements $\hat{\sigma}_x\hat{\sigma}_x$, $\hat{\sigma}_y\hat{\sigma}_y$ and $\hat{\sigma}_z\hat{\sigma}_z$. The measurement results for the three operators are shown in Fig. 4, each of which took about 60 hours. According to Eq. 6, we achieve an experimental fidelity of $0.65 \pm 0.03$.

As can be seen from the above experimental results, all the teleportation fidelities are well beyond the state estimation limit of $0.40$[14] for a two-qubit composite system, hence successfully demonstrating quantum teleportation of a two-qubit composite system. The imperfection of the fidelities is mainly due to the noise caused by emission of two pairs of down converted photons by a single source[6], which could produce four pairs in a single run and contribute to six-fold coincidence.

In summary we have developed and exploited a six-photon interferometer to report the first experimental demonstration of a two-qubit composite system. Not only does our experiment present an important step towards teleportation of a complex system, the techniques developed also enable future experimental investigations on novel quantum communication and computation protocols. For example, using a slightly modified experimental setup one can first prepare photons 3, 4, 5 and 6 in a four-photon cluster state[25] and further exploit the four-photon cluster state to demonstrate teleportation-based CNOT operation between photons 1 and 2, which is the kernel of fault-tolerant quantum computation[3]. In addition, whereas a modified six-photon interferometer can be used to prepare a six star-ring cluster state[26] and further implement universal quantum error-correction code[16] by performing a projection measurement on the input qubit, our six-photon interferometer can also be modified to produce various four, five and six-photon clusters[27] which are essential resources for one-way quantum computation[17,18].

**Acknowledgements**

This work was supported by the Marie Curie Excellent Grant of the EU and the Alexander von Humboldt Foundation. This work was also supported by the National Natural Science Foundation of China and the Chinese Academy of Sciences.

**Competing interests statement**

The authors declare that they have no competing financial interests.

**Correspondence** and requests for materials should be addressed to J.-W. Pan (jian-wei.pan@physi.uni-heidelberg.de).

**Figure Captions:**

**Figure 1**  Scheme showing principle of two-qubit quantum teleportation.

**Figure 2**  A schematic drawing of the experimental setup. The UV pulse passes through a $\beta$-barium borate (BBO) crystal to generate a polarization-entangled photon pair in modes 3 and 5 (i.e. the first ancillary entangled photon pair). After the first BBO, a 10cm-focus lens is introduced to refocus the UV pulse onto the second BBO to produce another entangled photon pair in modes 1 and 2 (to prepare the two qubits to be teleported). Reflected by a concave mirror, the UV pulse pumps once more into the second BBO and generates the third entangled photon pair in modes 4 and 6 (i.e. the second ancillary photon pair). Prism 1 and Prism 2 both mounted on step motors are used to compensate the time delay for the interference on polarizing beam-splitters PBS13 and PBS24, respectively. PBS5 and PBS6 are utilized to verify the teleported state with the help of wave plates in front of them. The photons are all detected by silicon avalanched single-photon detectors. Coincidences are recorded with a coincidence unit clocked by the infrared laser pulses.

**Figure 3** (A) Experimental results for $|\chi\rangle_A$ teleportation. (B) Experimental results for $|\chi\rangle_B$ teleportation. Each measurement takes 60 hours.

**Figure 4** Experimental results for $|\chi\rangle_C$ teleportation in the three complementary bases of $|H\rangle/|V\rangle$ (A), $|+\rangle/|-\rangle$ (B) and $|R\rangle/|L\rangle$ (C) corresponding to the three different local measurement $\hat{\sigma}_x\hat{\sigma}_x$, $\hat{\sigma}_y\hat{\sigma}_y$ and $\hat{\sigma}_z\hat{\sigma}_z$. Each measurement takes 60 hours.

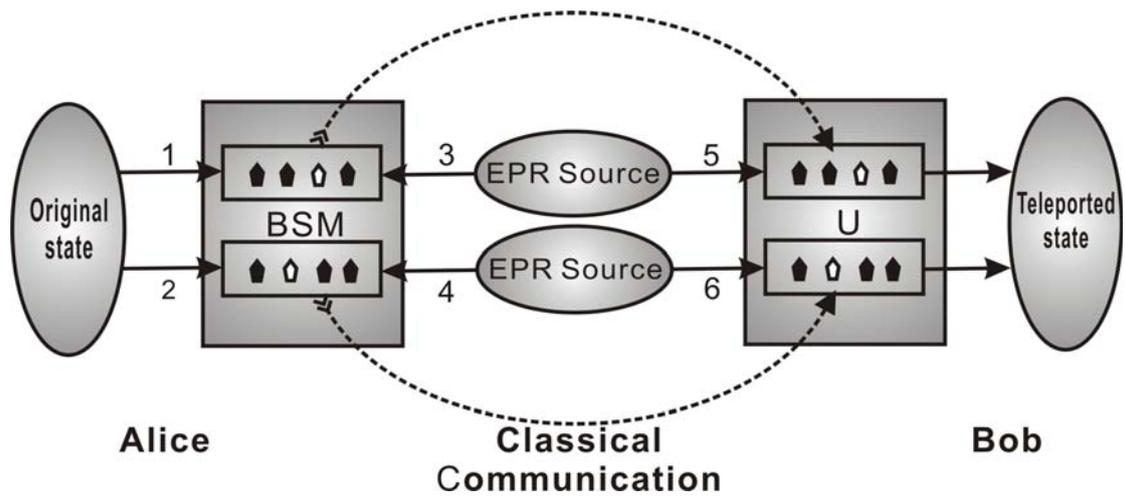

**Figure 1**

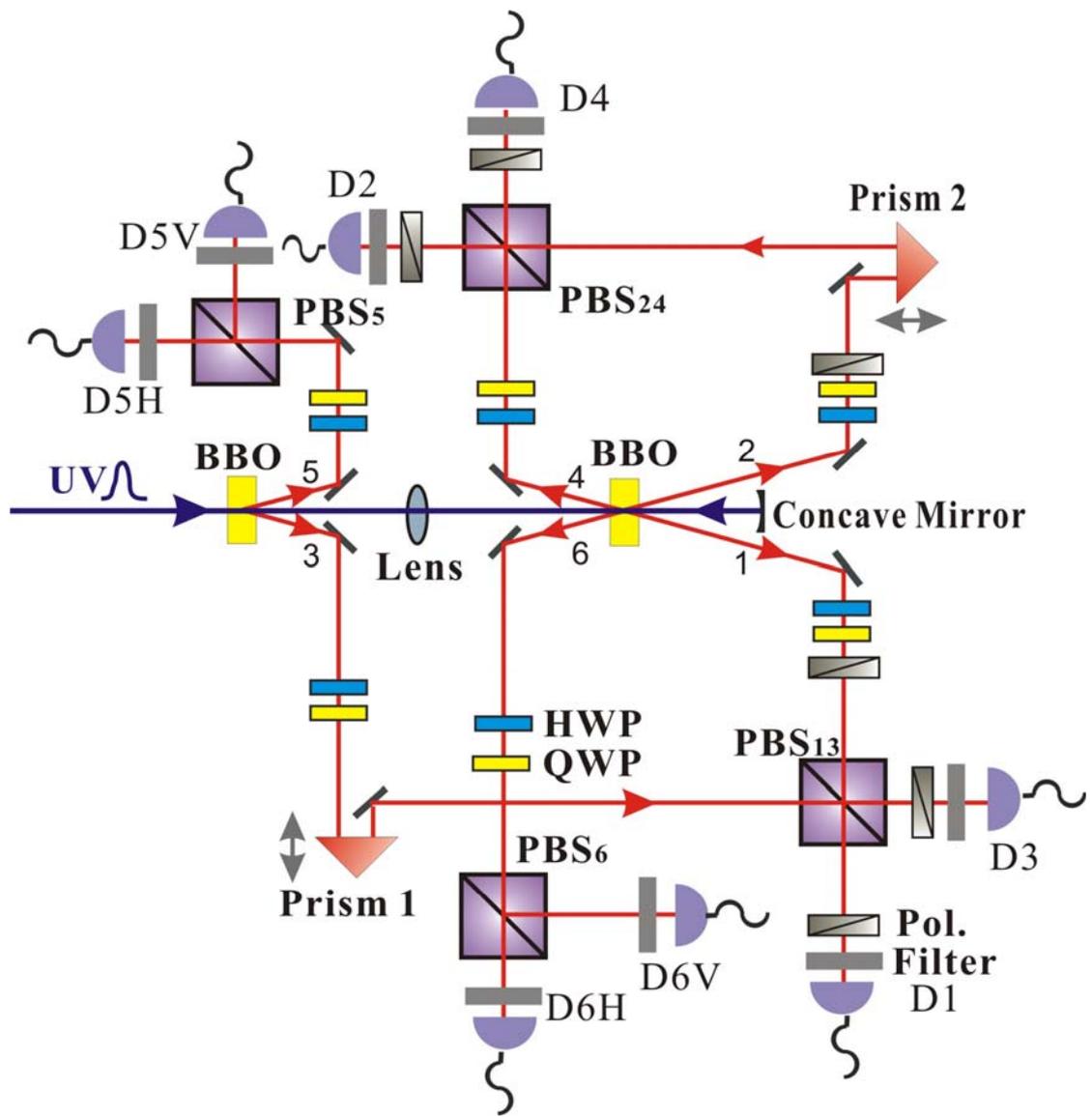

**Figure 2**

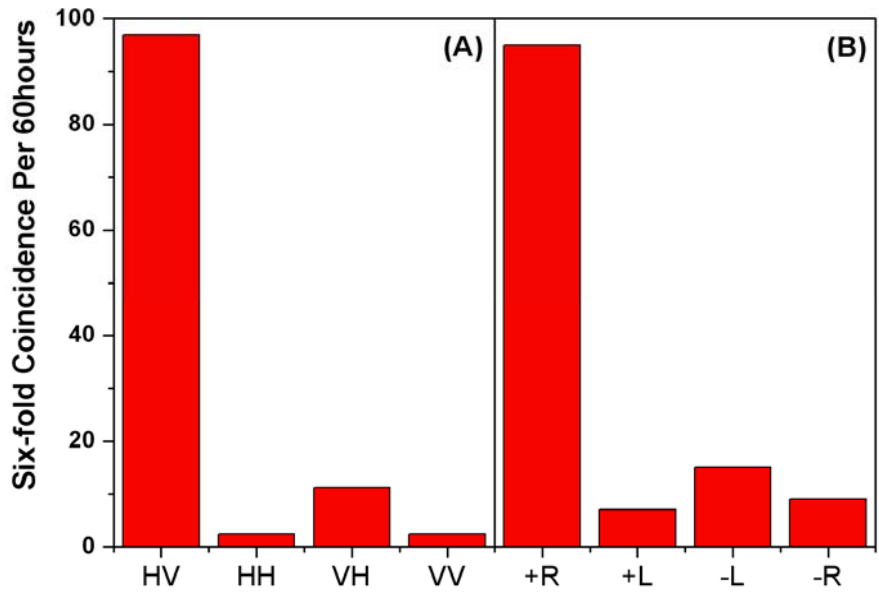

**Figure 3**

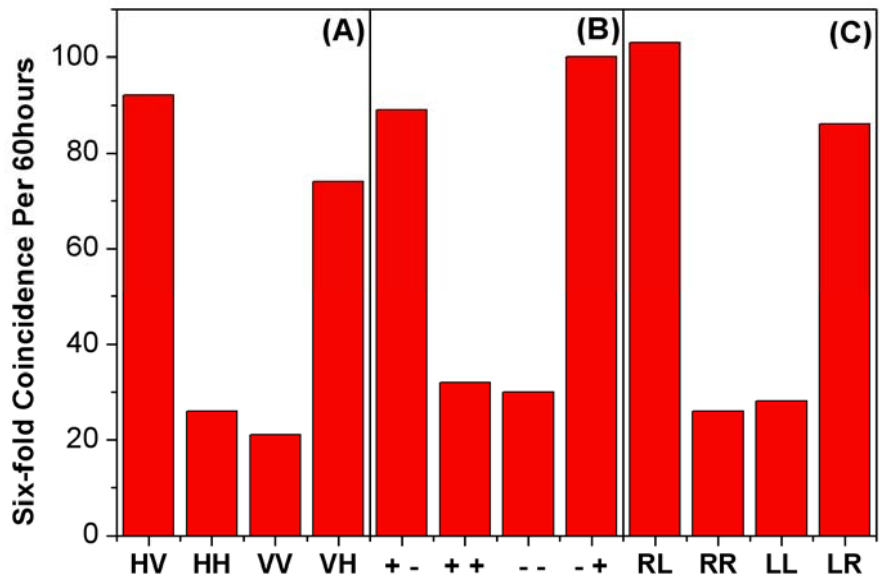

**Figure 4**